\documentclass[aps,preprint,nofootinbib,letterpaper]{revtex4}%
\usepackage{amssymb}
\usepackage{amsfonts}
\usepackage{amsmath}
\usepackage{amsmath}
\usepackage{graphicx}
\usepackage{silence} 
\usepackage{color}
\begin{document}
\title{Bumpy horizons from non-linear sigma models}
\author{Fabrizio Canfora$^1$, {Nicolás Grandi}$^2$\footnote{Corresponding author {\tt grandi@fisica.unlp.edu.ar}}, Carla Henríquez-Báez$^3$, Julio Oliva$^4$}
\affiliation{$^{1}$Centro de Estudios Científicos (CECS), Casilla 1469, Valdivia, Chile and Facultad de Ingeniería, Universidad San Sebastián, General Lagos 1163, Valdivia 5110693, Chile.}
\affiliation{$^{2}$Instituto de Física La Plata, CONICET and Departamento de Física Dr. Emil Bose, UNLP Casilla de correos 67, 1900 La Plata, Argentina. }
\affiliation{$^{3}$Centro Multidisciplinario de Física, Vicerrectoría de Investigación, Universidad Mayor, Camino La Pirámide 5750, Santiago, Chile.}
\affiliation{$^{4}$Departamento de F\'{\i}sica, Universidad de Concepci\'{o}n, Casilla 160-C, Concepci\'{o}n, Chile.\\~\\}
\date{March 31, 2026}
\begin{abstract}
We show that black hole horizons with non-constant curvature can be constructed in any dimension, when the source of the Einstein equations is given by the stress-energy tensor of a non-linear sigma model of a particular class.
%\newline 
%~
%\newline  
%\centerline{\em Essay written for the Gravity Research Foundation 2026 Awards for Essays on Gravitation.}
\end{abstract}
\maketitle

\section{Introduction}
Non-linear sigma models (NLSMs) represent a cornerstone in modern theoretical physics, providing highly effective frameworks for describing a wide array of physical phenomena. Historically, they have been crucial as low-energy effective field theories for pions in chiral perturbation theory  \cite{CHPT1, CHPT2, CHPT3, CHPT4}. In the context of general relativity, NLSMs arise naturally from the spacetime geometry via dimensional reduction; this correspondence facilitates the derivation of exact analytical solutions, as it maps the Einstein field equations onto an integrable system \cite{Belinski:2001ph}. Furthermore, in the framework of string theory, the NLSM provides a fundamental description of string propagation in curved backgrounds \cite{Green:2012oqa}. Simultaneously, they are ubiquitous in both gauged and ungauged supergravity theories, where they characterize the dynamics of the scalar sector 
\cite{Trigiante:2016mnt}.

Standard results in general relativity dictate that asymptotically flat, stationary black holes in vacuum must possess spherical horizon topologies \cite{Hawking:1971vc}. This constraint is relaxed in the presence of a negative cosmological constant, which allows for black holes with more complex, higher-genus horizon surfaces \cite{Lemos:1994xp,Mann:1996gj,Vanzo:1997gw,Brill:1997mf}. While such solutions are typically asymptotically locally AdS, their horizons are characterized by a constant local curvature. These geometries are derived as quotients of maximally symmetric spaces, a process that restricts the isometry group and can yield horizons that lack Killing vectors entirely despite their constant curvature.

Recently, black holes with ``bumpy'' horizons - characterized by a non-trivial dependence of the  curvature on the horizon coordinates - were shown to be obtained by coupling gravity to an SU(2) NLSM whose scalar fields satisfy a set of Cauchy-Riemann equations 
\cite{Canfora:2026col}. 
This framework was subsequently extended to more general NLSMs  \cite{Canfora:2026kwj} { from bottom-up approach}, showing that it is possible to construct a wide variety of exact solutions, including neutral, charged, and magnetized black holes with bumpy horizons, as well as black strings, $p$-branes, bumpy stars, and anisotropic cosmologies. These newly discovered solutions prove to be highly versatile and can be consistently embedded into supergravity frameworks.

In this work, we present a detailed construction of black holes with bumpy horizons in even dimensional Einstein gravity coupled to a non-linear sigma model, {following a top-down approach}. We start with a generic target space metric and a general ansatz for the metric and the scalar fields, and then apply the necessary restrictions to obtain a complete solution. 

% Specifically, we present black holes characterized by "bumpy" or irregular horizons.

%\newpage

\section{Action and equations of motion}
We consider an Einstein-Hilbert action for gravity with a cosmological constant in $n$ spacetime dimensions, coupled to a non-linear sigma model. The total action is
\begin{equation}
S =S_{\sf EH}+S_{\sf NLSM}= \int d^n x \, \sqrt{-g} \, \left( \frac{1}{2\kappa} (R - 2\Lambda) +  h_{i\bar j}(\varphi,\bar\varphi) \, g^{\mu\nu} \partial_\mu \varphi^i \partial_\nu \bar \varphi^{\bar j} \right),
\label{eq:action}
\end{equation}
where $\varphi^i$ are a set of complex scalar fields, and $h_{i\bar j}(\varphi,\bar\varphi)$ is a Hermitian metric in target space. In what follows, we set $\kappa=1$, since it can be re-absorved into $h_{i\bar j}(\varphi, \bar\varphi)$.

We adopt a static metric ansatz of the form 
\begin{equation}
ds^2 = -f(r) N^2(r)  dt^2 + \frac{dr^2}{f(r)} + r^2 \gamma_{a\bar b}(\zeta,\bar\zeta)\, d\zeta^a d\bar \zeta^b,
\label{eq:metric}
\end{equation}
where the transverse manifold is described as a complex space with a Hermitian metric $\gamma_{a\bar b}(\zeta,\bar\zeta)$. Notice that this ansatz implies that the total spacetime dimension is even. The transverse metric is taken to be independent of $t$ and $r$, and the sigma model fields are assumed to depend only on the transverse coordinates
\begin{equation}
\varphi^i=\varphi^i(\zeta,\bar\zeta)\,,
\qquad\qquad\qquad 
\bar\varphi^{\bar i}=\bar\varphi^{\bar i}(\zeta,\bar\zeta)\,.
\label{eq:scalars}
\end{equation}

With the ansatz above, the non-trivial Einstein equations for the system read %can be written down as
 \small
\begin{align}
&(tt)~~~~-\!\frac{f N^2}{2r^2} \left( R^{(\gamma)}\! - (n\!-\!2)(n\!-\!3) f - (n\!-\!2) r f'-2r^2\Lambda  \right) =
 -\frac{fN^2}{2%WARNING
 r^2} \,  h_{i\bar j}
 \gamma^{\bar b a}\!\left(\partial_a\varphi^i\partial_{\bar b}\bar\varphi^{\bar j}
 +\partial_{\bar b}\varphi^i\partial_{a}\bar\varphi^{\bar j}
 \right), \nonumber\\ 
&(rr)~~~~\frac{(n\!-\!2) N'}{r N} - \frac{1}{2f r^2} \left( R^{(\gamma)} - (n\!-\!2)(n\!-\!3) f -(n\!-\!2) rf'-2r^2\Lambda \right)
=
\nonumber\\ &\qquad\qquad\qquad\qquad\qquad\quad\qquad\qquad\qquad\qquad\qquad\qquad\qquad\ 
=-\frac{1}{2%WARNING
r^2f} \,  h_{i\bar j}\gamma^{\bar b a}\left( \partial_a\varphi^i\partial_{\bar b}\bar\varphi^{\bar j}
 + \partial_{\bar b}\varphi^i\partial_{a}\bar\varphi^{\bar j}
 \right),\nonumber
\\ 
&(a\bar b)~~~~G^{(\gamma)}_{a\bar b} + \gamma_{a\bar b} r^2\left(  \frac{1}{2} f''  + \frac{(n\!-\!3)}r  f'  + \frac{(n\!-\!3)(n\!-\!4)}{2r^2} f + \frac{3}{2}  \frac{f' N'}{N} +  \frac{f N''}{N}   + (n\!-\!3)   \frac{fN'}{rN}+\Lambda \right)
=\nonumber\\&
\qquad\qquad\qquad\qquad\ \
\quad\quad\qquad 
=
%2
h_{i\bar j} \left(\partial_a \varphi^i \partial_{\bar b}\bar \varphi^{\bar j}+
 \partial_{\bar b} \varphi^i \partial_a\bar \varphi^{\bar j}
 \right)-\frac12%WARNING
\gamma_{a\bar b}  \gamma^{\bar dc} h_{i\bar j} \,\left( \partial_c\varphi^i\partial_{\bar d}\bar\varphi^{\bar j}
 + \partial_{\bar d}\varphi^i\partial_{c}\bar\varphi^{\bar j}
 \right),
 \nonumber\\ 
 &(ab)~~~~G_{ab}=%-2
 h_{i\bar j} \left(\partial_a \varphi^i \partial_{ b}\bar \varphi^{\bar j}+
 \partial_{ b} \varphi^i \partial_a\bar \varphi^{\bar j}
 \right),
 \nonumber\\
 &(\bar a \bar b)~~~~G_{\bar a\bar b}=  %-2
 h_{i\bar j} \left(\partial_{\bar a} \varphi^i \partial_{\bar b}\bar \varphi^{\bar j}+
 \partial_{\bar b} \varphi^i \partial_{\bar a}\bar \varphi^{\bar j}
 \right),
 \label{eq:Einstein.equations}
\end{align}
\normalsize
where $R^{(\gamma)}$ is the Ricci scalar of the transverse metric $\gamma_{a\bar b}$ and $G^{(\gamma)}_{a\bar b}$ is its Einstein tensor.  These equations have to be solved in combination with the equations of motion of the scalar fields, which read 
\begin{equation}
    \frac1{\sqrt{\gamma}}\partial_a\left(\sqrt{\gamma}\gamma^{\bar b a}h_{i\bar j}\partial_{\bar b}\varphi^i\right)+    
    \frac1{\sqrt{\gamma}}\partial_{\bar b}\left(\sqrt{\gamma}\gamma^{\bar b a}h_{i\bar j}\partial_{a}\varphi^i\right)=
    h_{i\bar k,\bar j}\,\gamma^{\bar b a}\!\left(\partial_a\varphi^i\partial_{\bar b}\bar\varphi^{\bar k}+\partial_a\bar\varphi^{\bar k}\partial_{\bar b}\varphi^i\right)
%    \qquad\qquad\mbox{(and its complex conjugate)}    
\label{eq:scalar.equation}
\end{equation}
where $\gamma$ is the determinant of the transverse metric.

\section{Solution of the equations}
\label{sec:solution}
In the Einstein equations \eqref{eq:Einstein.equations} we have made no simplifications at all, keeping the geometry on the right-hand side and the matter on the left-hand side. However, the $(tt)$ equation can be rewritten in the much simpler form, 
\begin{equation}
 (n-2) r f' +(n-2)(n-3) f+2r^2\Lambda =
 R^{(\gamma)}-2 
  h_{i\bar j}
 \gamma^{\bar ba}\left(\partial_a\varphi^i\partial_{\bar b}\bar\varphi^{\bar j}
 + \partial_{\bar b}\varphi^i\partial_{a}\bar\varphi^{\bar j}
 \right),
\end{equation}
In this expression, the left-hand side depends on the radial coordinate $r$ only, while the right-hand side depends on the transverse variables $\zeta,\bar\zeta$. This allows for a separation-of-variables ansatz:
\begin{align}
& (n-2) r f' +(n-2)(n-3) f+2r^2\Lambda  =\alpha,
\label{eq:separation.of.variables.f}
%\nonumber 
\\ & 
%\qquad\quad
R^{(\gamma)}- 
 h_{i\bar j}
 \gamma^{\bar ba}\left(\partial_a\varphi^i\partial_{\bar b}\bar\varphi^{\bar j}
 +\partial_{\bar b}\varphi^i\partial_{a}\bar\varphi^{\bar j}
 \right)
 =\alpha,
\label{eq:separation.of.variables.gamma}
\end{align}
where $\alpha$ is a separation constant. 
The equation \eqref{eq:separation.of.variables.f} can be solved straightforwardly for $f(r)$, resulting in the standard form of the blackening factor for a black hole in $n$ spacetime dimensions
\begin{equation}
f(r) = \frac{\alpha}{(n-2)(n-3)} -\frac{2M}{r^{n-3}}  - \frac{2\Lambda}{(n-2)(n-1)} r^2 
\label{eq:blackening.factor}
\end{equation}
Here $M$ is an integration constant representing the mass of the configuration. This yields a geometry that possesses an event horizon at $r_h$ where $f(r_h)=0$. The transverse metric $\gamma_{a\bar b}(\zeta,\bar\zeta)$ is then re-interpreted as the horizon metric. 

On the other hand, when the ansatz \eqref{eq:separation.of.variables.gamma} is substituted into the $(rr)$ equation, we obtain $N'=0$, implying that $N$ is a constant that can be fixed to $N=1$ by a rescaling of the time variable $t$. Notice that this result, which is standard in the case of a vacuum black hole, is obtained here in the presence of scalar matter sources.

Once the explicit forms for $f$ and $N$ have been substituted into \eqref{eq:Einstein.equations}, the remaining Einstein equations simplify to
\small
\begin{align}
&(a\bar b)~~~~ G^{(\gamma)}_{a\bar b} + \gamma_{a\bar b} \frac{(n-4)\alpha}{2(n-2)}=%2
h_{i\bar j} \left(\partial_a \varphi^i \partial_{\bar b}\bar \varphi^{\bar j}+
 \partial_{\bar b} \varphi^i \partial_a\bar \varphi^{\bar j}
 \right)-\frac12
\gamma_{a\bar b}  \gamma^{\bar dc} h_{i\bar j} \,\left(\partial_c\varphi^i\partial_{\bar d}\bar\varphi^{\bar j}
 +\partial_{\bar d}\varphi^i\partial_{c}\bar\varphi^{\bar j}
 \right),\nonumber
 \\ 
 &(a b)~~~~G_{ab}=- %2
 h_{i\bar j} \left(\partial_a \varphi^i \partial_{ b}\bar \varphi^{\bar j}+
 \partial_{ b} \varphi^i \partial_a\bar \varphi^{\bar j}
 \right),\nonumber
 \\
 &(\bar a\bar b)~~~~G_{\bar a\bar b}= - %2
 h_{i\bar j} \left(\partial_{\bar a} \varphi^i \partial_{\bar b}\bar \varphi^{\bar j}+
 \partial_{\bar b} \varphi^i \partial_{\bar a}\bar \varphi^{\bar j}
 \right),
 \label{eq:Einstein.remaining}
\end{align}
\normalsize
Taking the trace of the $(a\bar b)$ equation with the horizon metric $\gamma^{\bar b a}$ recovers equation \eqref{eq:separation.of.variables.gamma}, confirming consistency. To solve the $(ab)$ and $(\bar a\bar b)$ equations, we first choose the horizon metric to be Kähler, which immediately gives $G_{ab}=G_{\bar a\bar b}=0$. The equations then impose the constraints $\varphi^i=\varphi^i(\zeta)$ and $\bar\varphi^{\bar i}=\bar\varphi^{\bar i}(\bar \zeta)$. That is, the scalar fields are holomorphic functions of the horizon coordinates, while their complex conjugates are antiholomorphic.

Up to this point, we are left with an equation for the horizon metric $\gamma_{a\bar b}$, given by 
\begin{align}
& G^{(\gamma)}_{a\bar b} + \gamma_{a\bar b} \frac{(n-4)\alpha}{2(n-2)}=%2
h_{i\bar j} \partial_a \varphi^i \partial_{\bar b}\bar \varphi^{\bar j} -\frac12%WARNING
\gamma_{a\bar b}   h_{i\bar j} \,
\gamma^{\bar d c}\partial_c\varphi^i\partial_{\bar d}\bar\varphi^{\bar j},
\label{eq:G.ab}
\end{align}
and the equations \eqref{eq:scalar.equation} for the scalar fields
\begin{equation} 
    \frac1{\sqrt{\gamma}}\partial_{\bar b}\left(\sqrt{\gamma}\gamma^{\bar b a}h_{i\bar j}\right)\partial_{a}\varphi^i=
    h_{i\bar k,\bar j}\,\gamma^{a\bar b}\partial_a\varphi^i\partial_{\bar b}\bar\varphi^{\bar k} 
    \label{eq:scalar.equation.holomorphic}
\end{equation}

Because we assumed that the horizon metric $\gamma_{a\bar b}(\zeta,\bar\zeta)$ is Kähler, it can be obtained from a Kähler potential $k(\zeta,\bar\zeta)$ through the formula $\gamma_{a\bar b}(\zeta,\bar\zeta)=k_{,a\bar b}(\zeta,\bar\zeta)$. Now, to simplify the scalar equations \eqref{eq:scalar.equation.holomorphic}, we assume that the horizon metric is diagonal in the complex coordinates $\zeta^a$, namely $\gamma_{a\bar b}(\zeta,\bar\zeta)=0$ whenever $a\neq b$. In terms of the Kähler potential $k(\zeta,\bar\zeta)$, this implies that it is a sum of independent terms, each depending on one of the $(\zeta^a,\bar\zeta^{\bar a})$ pairs
\begin{equation}
    k(\zeta,\bar\zeta)=\sum_ak^{(a)}(\zeta^a,\bar\zeta^{\bar a})
\end{equation}
Then the metric is also a sum of terms depending on each coordinate pair, namely
\begin{equation}
    ds^2=\gamma_{a\bar b}(\zeta,\bar\zeta) d\zeta^ad\bar\zeta^{\bar b}=\sum_a e^{P_{(a)}(\zeta^a,\bar\zeta^{\bar a})}\,d\zeta^ad\bar \zeta^{\bar a}
\end{equation}
where the conformal factor $e^{P_{(a)}(\zeta^a,\bar\zeta^{\bar a})}=k^{(a)}_{,a\bar a}(\zeta^a,\bar\zeta^{\bar a})$ is defined in terms of the Kähler potential $k^{(a)}(\zeta^a,\bar\zeta^{\bar a})$.
With these assumptions, the operator acting on the scalar fields in equation \eqref{eq:scalar.equation.holomorphic} separates into a sum of terms depending on each coordinate pair $(\zeta^a,\bar\zeta^{\bar a})$. Now, since each block is two-dimensional, the horizon metric completely disappears from these equations, leading to
\begin{equation} 
    %\sum_a\partial_{\bar a} h_{i\bar j}\, \partial_{a}\varphi^i= 
    h_{i\bar j,\bar k} \sum_a\partial_{a}\varphi^i \partial_{\bar a}\bar \varphi^{\bar k}=h_{i\bar k,\bar j}\sum_a \partial_a\varphi^i\partial_{\bar a}\bar\varphi^{\bar k} 
    \label{eq:scalar.equation.holomorphic.2}
\end{equation}
If we assume that the target space metric is also Kähler, then it satisfies $h_{i\bar j,\bar k} =h_{i\bar k,\bar j}$ and the scalar equation is automatically satisfied. The corresponding Kähler potential satisfies $h_{i\bar j}(\varphi,\bar\varphi)= K_{,i\bar j}(\varphi,\bar\varphi)$. 

With all the above, the only remaining equations are \eqref{eq:G.ab}, which now read
\begin{align}
&\partial_a\partial_{\bar b}(%2
K+\log\det k_{,e\bar f}) -\frac12k_{,a\bar b} \left(\frac{(n-4)\alpha}{n-2}+\gamma^{\bar d c}\partial_c\partial_{\bar d}(K+\log\det k_{,e\bar f})\right)=0
\label{eq:G.ab.2}
\end{align}
Let us first analyze the off-diagonal components $a\neq b$ of this expression $\partial_a\partial_{\bar b}(2K-\log\det k_{,e\bar f}) =0$. They imply that the combination $K+\log\det k_{,e\bar f}$ is a sum of different terms, each of which depends on a different coordinate pair $(\zeta^a,\bar\zeta^{\bar a})$. Since $\log\det k_{,e\bar f}=2\sum_a P_{(a)}(\zeta^a,\bar\zeta^{\bar a})$, this is automatically satisfied by the second term. Regarding the first term, this condition forces us to restrict our NLSM to a Kähler potential with the form
\begin{equation}
    K(\varphi,\bar\varphi)=\sum_i K_{(i)}(\varphi^i,\bar\varphi^{\bar i})
\end{equation}
and then impose that each of the fields $\varphi^i(\zeta)$ depends on a single complex coordinate $\zeta^a$. The diagonal components then read
\begin{align}
&-\sum_{c\neq a} e^{-P_{(c)}}\partial_c\partial_{\bar c}\!\left(K+ 2P_{(c)}\right) +e^{-P_{(a)}}\partial_a\partial_{\bar a} \left(K+ 2P_{(a)}\right)=\frac{(n-4)\alpha}{2(n-2)}
\label{eq:G.ab.3}
\end{align}
In the left-hand side of these equations, each term depends on a different complex coordinate $\zeta^\alpha$. This implies that we can make a separation of variables, to obtain
\begin{align}
&e^{-P_{(a)}}\partial_a\partial_{\bar a} \left(K_{(a)}+ 2P_{(a)}\right)=-\frac{(n-4)\alpha_a}{2(n-2)}
\qquad\qquad\qquad\qquad
\sum_{c\neq a}\alpha_c-\alpha_a=\alpha 
\label{eq:separation.Liouville}
\end{align}
where the $\alpha_a$ are the separation constants, and we assumed that the field $\varphi^a$ depends on the coordinate $\zeta^a$. However, the constraint between them given by the second equation can only be solved if all the $\alpha_a$ are  equal $\alpha_a=2\alpha/(n-4)$. This leaves us with
\begin{align}
&\partial_a\partial_{\bar a} \left(K_{(a)}+2P_{(a)}\right)=-\frac{\alpha}{(n-2)}e^{P_{(a)}}
\label{eq:Liouville}
\end{align}
There are a set of non-homogeneous Liouville equations, where the source is given by the derivatives of the target Kähler potential. For any given set of holomorphic $\varphi^{(a)}(\zeta^a)$, a complete solution of our problem is obtained by solving equations \eqref{eq:Liouville}.

\newpage
\section{Summary and discussion}
\label{sec:discussion}

Let us summarize our result: when the action \eqref{eq:action} has a Kähler target-space metric with a block-diagonal form, its solutions are obtained using the ansatz \eqref{eq:metric} and \eqref{eq:scalars}. They are black holes whose blackening factor takes the standard higher-dimensional form \eqref{eq:blackening.factor}, while the horizon metric is a block-diagonal Kähler metric that solves \eqref{eq:Liouville}. Finally, the scalar fields are holomorphic functions of a single horizon coordinate.

The first observation is that the horizon curvature is not a constant, resulting in a ``bumpy'' horizon geometry. This can be seen in equation \eqref{eq:Liouville}, which implies for the scalar curvature
\begin{equation}
    R^{(\gamma)}=-2\sum_ a e^{-P_{(a)}}\partial_a\partial_{\bar a}
    P_{(a)}= 2\alpha +2\sum_ a e^{-P_{(a)}}\partial_a\partial_{\bar a}K_{(a)}
    \label{eq:horizon.curvature}
\end{equation}
In the right-hand side, the target Kähler potential $K_{(a)}(\varphi^a,\bar\varphi^{\bar a})$ is a function of the field $\varphi^a$ (and its complex conjugate $\bar\varphi^{\bar a}$) which is an arbitrary holomorphic (respectively anti-holomorphic) function of the coordinate $\zeta^a$.

The conclusion is that the diversity of horizon curvatures is as large as that of holomorphic functions. This strongly restricts the case of a compact horizon, since the only holomorphic function on a compact manifold is the constant function. In the non-compact case, the set of holomorphic functions is infinite, implying infinitely many different bumpy black holes.

{

Another interesting point is that the condition to have holomorphic fields saturates a Bogomol'nyi-Prasad-Sommerfield bound on the scalar field energy, which can be written when the geometry is specialized to our metric ansatz \cite{Canfora:2026col, Canfora:2026kwj}.

As mentioned in the introduction, the present construction { was originally presented in \cite{Canfora:2026col} for the particular case of a two-dimensional horizon and an SU(2) NLSM. Rather than imposing Killing symmetries in the usual sense, we required the metric to be compatible with the validity of the first-order BPS equations for the scalar field. In this sense, our approach is closer in spirit to Carter's search for metrics compatible with separability of the Klein–Gordon equation \cite{Carter} than to standard symmetry-based constructions.

In \cite{Canfora:2026kwj}, we proved that it } can be generalized to include a wide range of additional matter sources, to odd-dimensional geometries, and even to time-dependent solutions. { The main new contribution of the present work is that we re-derive the solution to the purely Einstein-NLSM model case, without making any {\em a priori} assumption other than the ansatz \eqref{eq:metric}-\eqref{eq:scalars}. At each step in the derivation, we introduce the necessary further restrictions on the model and the ansatz in order for the separation procedure to work.

For example, in passing from equation \eqref{eq:Einstein.remaining} to \eqref{eq:G.ab} and \eqref{eq:scalar.equation.holomorphic}, we restricted the horizon metric to be Kähler, which implies that the scalar fields are holomorphic. Had we not taken that step, the procedure would have become stuck and the equations would not have been separated. Later, the assumption that the horizon metric is block-diagonal led us to restrict the model to a Kähler target space, in equation \eqref{eq:scalar.equation.holomorphic.2}. The resulting equation, \eqref{eq:G.ab.2}, could only be solved if the target space is also block-diagonal. As a final step, we separated variables and proved that all the resulting equations contain the same value of the separation constant, resulting in \eqref{eq:Liouville}.

The above considerations lead to the main conclusion of the present paper: the ansatz for the metric and the scalar fields originally proposed in \cite{Canfora:2026col} and generalized in \cite{Canfora:2026kwj} cannot be extended to non-product geometries for either the horizon or the target space, and it reduces to a single non-homogeneous Liouville equation that is identical for each two-dimensional factor of the horizon metric.

As possible extensions of the present construction, an interesting possibility not considered in [14] that we would like to explore in a forthcoming publication is to couple the electromagnetic field both to gravity and to the SU(2) NLSM in 3+1 dimensions. In this case, the NLSM describes the Pions dynamics. In the non-gravitational case, we know that such coupling spoils the first order BPS equations, which were crucial in our manuscript as the scalar field become analytic.

As far as stability is concerned, a rigorous analysis requires a twofold path. First, we could try to embed our model into a suitable Supergravity theory, such that the resulting Bogomol'nyi-Prasad-Sommerfield sector contains the present solutions, which would immediately imply they are stable. A different path is the numerical study of the full linearized system. Regardless of the outcome, any decay must preserve the topological charge, which in our case corresponds to the two-dimensional vorticity. This implies that an unstable solution would decay into another configuration of equal vorticity, a class of configurations we are actively trying to construct.

}}

\newpage

\section*{Acknowledgments}
We thank Andrés Anabalón for comments and Andrés Gomberoff and Aldo Vera for early work on this topic. F. C. has been funded by FONDECYT Grants
No. 1240048, 1240043 and 1240247 and
also supported by Grant ANID EXPLORACIÓN 13250014.  C. H appreciates the support of FONDECYT postdoctoral Grant 3240632. J.O. was supported by FONDECYT grant 1221504. N.G. is partially supported by CONICET grant PIP-2023-11220220100262CO, and
UNLP grant 2022-11/X931. N.  G. thanks Concepción U., San Sebastián U.,  Mayor U., and CECS for hospitality during the early stage of this work.

\end{document}